# Dynamics on the Double Morse Potential: A Paradigm for Roaming Reactions with no Saddle Points

Barry K. Carpenter,[†] Gregory S. Ezra,[‡] Stavros C. Farantos,[§] Zeb C. Kramer,[∥] Stephen Wiggins*[⊥]

[†] School of Chemistry, Cardiff University, Cardiff CF10 3AT, United Kingdom

[‡] Department of Chemistry and Chemical Biology, Cornell University, Ithaca, NY 14853-1301, United States

[§] Institute of Electronic Structure and Laser, Foundation for Research and Technology – Hellas, and Department of Chemistry, University of Crete, Iraklion 711 10, Greece

[∥] Department of Chemistry and Biochemistry, La Salle University, 1900 West Olney Avenue, Philadelphia, PA 19141, United States

[⊥] School of Mathematics, University of Bristol, Bristol BS8 1TW, United Kingdom

**Abstract**: In this paper we analyze a two degree of freedom Hamiltonian system constructed from two planar Morse potentials. The resulting potential energy surface has two potential wells surrounded by an unbounded flat region containing no critical points. In addition, the model has an index one saddle between the potential wells. We study the dynamical mechanisms underlying transport between the two potential wells, with emphasis on the role of the flat region surrounding the wells. The model allows us to probe many of the features of the "roaming mechanism" whose reaction dynamics are of current interest in the chemistry community.

**Introduction**: In the 1978 English version of his famous book *Mathematical Methods of Classical Mechanics* V. I. Arnold made the provocative statement:

*"Analyzing a general potential system with two degrees of freedom is beyond the capability of modern science."*

Despite the great progress in our understanding on nonlinear dynamical systems theory since that time, this statement is still mostly true.

In this paper, we describe a new class of two-dimensional (2*d*) potential energy surface (PES) constructed from two planar Morse potentials whose rich dynamics certainly bears out Arnold's assessment of the situation. The PES has two potential wells, separated by an index one saddle point, and surrounded by an unbounded flat region containing no critical points. We are interested in the *fate* of trajectories that leave one of the potential wells. It is observed that such trajectories have three possible fates. They can 1) exit a potential well and re-enter the same potential well at a later time; 2) they can exit a potential well and enter the other potential well, or 3) they can exit a potential well, enter the flat region, and become unbounded. The collective dynamical behavior concerning inter-well transport embodied in this system has many of the features of the "roaming mechanism" for reaction dynamics that is of current interest in the chemistry community [1,2].

Roaming is a recently discovered mechanism for chemical reaction, i.e. breaking and/or forming chemical bonds between atoms [1]. Many studies of the roaming phenomenon in a variety of diverse settings for chemical reactions have been carried out in recent years. It is outside the scope of this paper to review the topic of roaming. However, in recent years there have been a number of reviews on the topic. See, for example, [2].

Briefly, we describe features of the roaming mechanism that are relevant to our model. Dynamically, roaming is a phenomenon in which an energized molecule, seemingly about to undergo dissociation into two fragments, instead exhibits behavior in which one fragment begins to orbit about the other one (typically occurring in a flat region of the PES), culminating in a re-encounter of the two fragments in some "reactive event." In our model "dissociation" corresponds to a trajectory leaving a well and becoming unbounded. "Recombination events" involve the potential wells, and the flat region of the PES is the "no man's land" wherein trajectories "decide" between a particular recombination event or dissociation. Since this region is flat, and contains no saddle points there must be some form of *dynamical* selection mechanism in operation. Understanding that mechanism is a focus of this paper. Indeed, we will see that the notion of a normally hyperbolic invariant manifold (NHIM) lies at the heart of this dynamical selection mechanism [3]. In the case of 2 degree of freedom (DoF) Hamiltonian systems the NHIM is an unstable periodic orbit (PO) [4].

**The Potential:** The 2 DoF Hamiltonian system that we consider consists of two identical planar Morse potentials [5] separated by a distance $2b$, as seen in Figures 1 and 2. In Cartesian coordinates this has the form:

$$V_{DM} = D_e \left(1 - e^{-\sqrt{\frac{k}{D_e}}\left(\sqrt{(x-b)^2+y^2}-r_e\right)}\right)^2 + D_e \left(1 - e^{-\sqrt{\frac{k}{D_e}}\left(\sqrt{(x+b)^2+y^2}-r_e\right)}\right)^2 - D_e \quad (1)$$

The parameter $D_e$ sets the energy of the asymptote as $x^2 + y^2 \to \infty$. The parameter $k$ determines the curvature of potential around the minima, whose positions are set by the parameters $b$ and $r_e$. For the studies carried out in this paper, $D_e = 100, k = 200, r_e = 1$. The parameter $b$ determines the distance between the two Morse functions. Taking the units of energy to be kcal mol$^{-1}$, and units of distance to be Å, the parameters of the single Morse potential functions in Eq. 1 correspond roughly to the vibration of a CH bond.

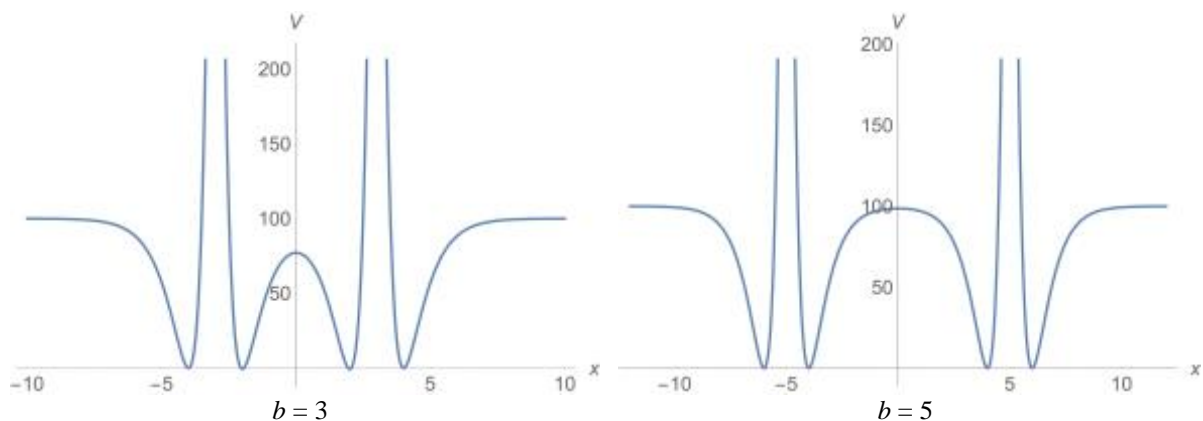

**Figure 1**. Sections through $V_{DM}$ at $y = 0$, for $b = 3$ and $b = 5$.

There is an index one saddle[1] at the origin. It has an energy below that of the asymptote, with the value being determined by the value of *b*. As *b* gets smaller, the index one saddle occurs at lower energy. This is the only saddle on the PES. In Figs. 1, 2, and 3, we illustrate different representations of the PES for *b* = 3 and *b* = 5. The "spikes" visible in the center of the two potential wells in Figs. 1 and 3 are a result of the "repulsive core" of the PES. In Fig. 3 the flat region of the PES surrounding the potential wells is readily apparent. The dynamics of trajectories in this region will be a focus of our studies.

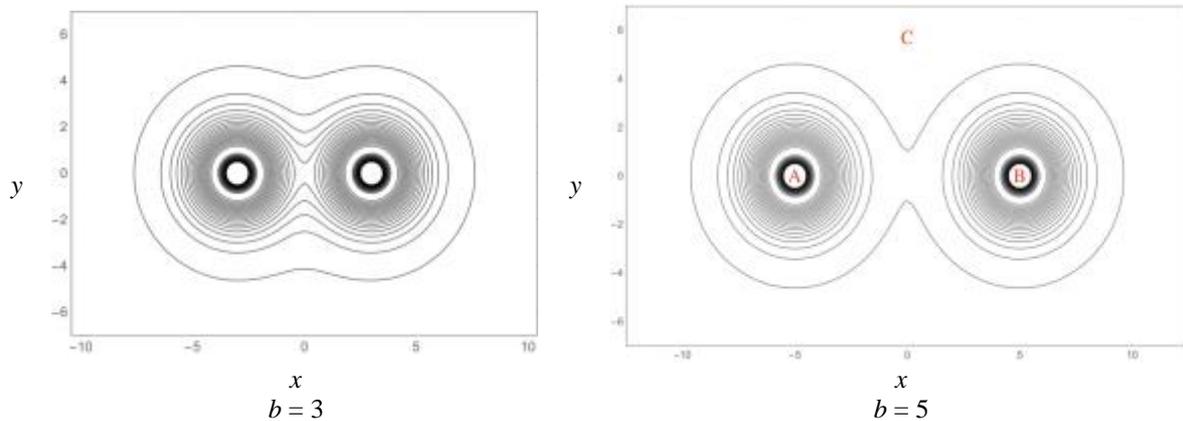

*b* = 3        *b* = 5

**Figure 2.** Contours of $V_{DM}$ for *b* = 3 and *b* = 5. The regions A, B and C of the potential, referred to in the text, are labeled in the right-hand panel.

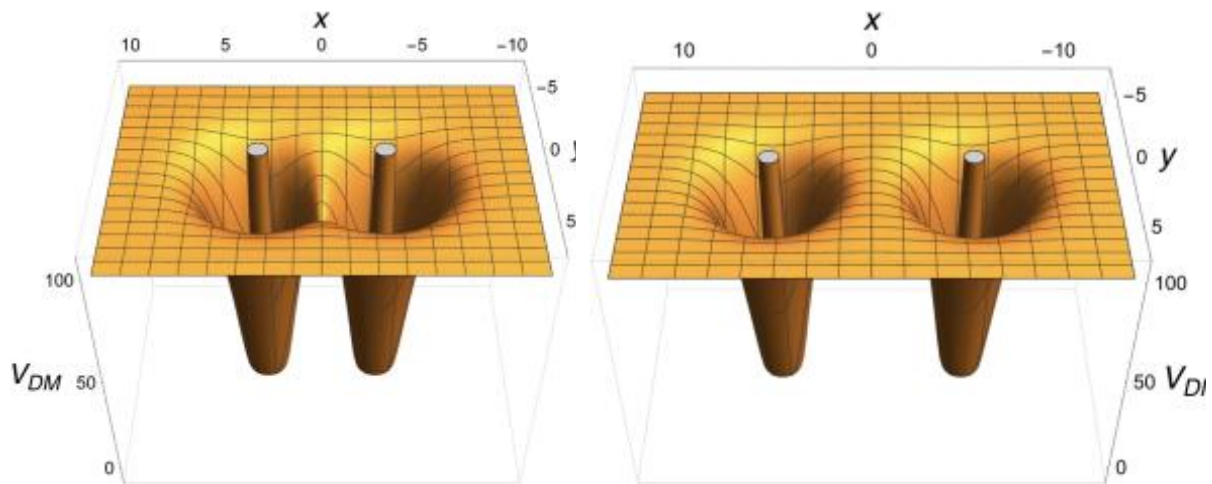

**Figure 3.** Graphical representations of Eq. 1, with parameter values $D_e$ =100, $k$ =200, $r_e$ =1. The left-hand panel has *b*=3, and the right-hand panel has *b*=5.

**Dynamics on the Double-Morse Potential:** Chemically, this potential could be considered to represent an atom interacting with a diatomic molecule, or, mechanically, a rigid dumbbell interacting with a particle. The equilibrium bond length between the two atoms of the diatomic is set by *b* and the bond length of the third atom to one of the atoms of the diatomic is set by $r_e$. In all the calculations described below, the kinetic energy, *T*, takes the form shown in Eq. 2.

---

[1] In the dynamical systems literature, for 2 DoF Hamiltonian systems, index one saddle points are referred to as equilibria of saddle-center stability type.

$$T = \frac{p_x^2}{2m} + \frac{p_y^2}{2m} \qquad (2)$$

*m* was set to a value of 1, and the Hamiltonian *H*, i.e. the sum of the kinetic and potential energy, is represented by Eq. 3.

$$H(p_x, p_y, x, y) = T(p_x, p_y) + V_{DM}(x, y) = E \qquad (3)$$

This is a conservative Hamiltonian and its value corresponds to the total energy of the system, E.

The dynamical phenomena that will be of principal concern in this paper are exhibited by trajectories that start and end in a well and, in the process, make an excursion through the flat region of the PES. One can indeed find such trajectories, as illustrated in Fig. 4 for a total energy of 100.001 on the PES with b = 3. Such trajectories may be identified as "roaming trajectories" of the type discussed in the chemistry literature [1,2].

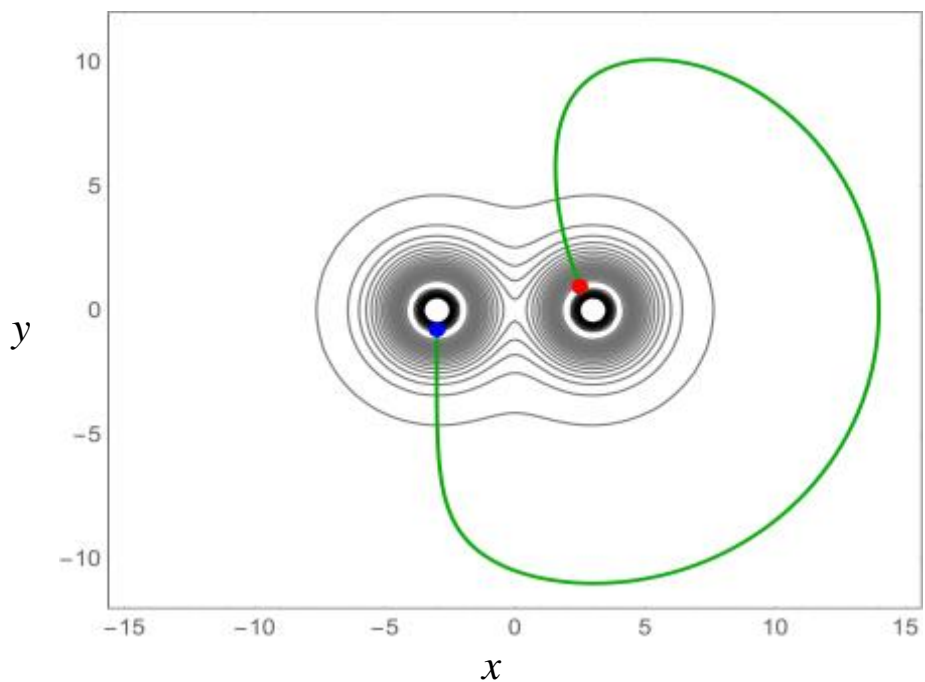

**Figure 4**. Illustration of a "roaming" trajectory at a total energy of 100.001 on the potential with *b* = 3. The blue dot indicates the beginning of the trajectory, and the red dot its end.

In the example shown in Fig. 4, the "roaming" trajectory corresponds to a situation in which the third atom migrates from one atom of the diatomic to the other. The interesting feature is that once the trajectory enters the flat region, it "turns around" and enters the other well, despite the fact there is no PES structure that might be the cause of such dynamics, e.g. a "roaming saddle". Here, the word "dynamics" is the key. Attempts to understand this phenomenon solely in terms of the structure of the PES, i.e. a function defined on configuration space, are doomed to failure. Understanding of the mechanism necessitates a phase space approach.

In order to characterize the behavior of the trajectories of interest three regions of configuration space are defined, and we will be concerned with the dynamical properties of trajectories that visit these three regions (with initial conditions also taken in one of the regions). The three regions are denoted by A, B, and C in Fig. 2. Regions A and B correspond to the two wells of the PES and C denotes the region where trajectories enter the flat part of the potential and become unbounded. Below we will give a precise dynamical definition of these regions. However, first we will consider an important, and general, dynamical property of this 2 DoF Hamiltonian system.

**Chaos and the Non-Existence of a Second Constant of Motion:** The Morse potential defines a central force, and the double Morse potential is a type of "two-center problem". The two-center problem where the two potentials are of either Coulomb or Kepler type has received a great deal of attention in the dynamics community, both because of its physical relevance and because it is completely integrable. An excellent review of the two-center problem, among other related topics has been given in [6].

While it is tempting to seek an additional constant of the motion for the double Morse potential, we provide numerical evidence, which (in combination with a theorem due to Moser [7]) rules out the existence of a second (analytic) integral for the double Morse potential.

A Poincaré surface of section for a trajectory started at the index one saddle of the $b = 5$ PES and recorded for passage through $x = 0$ in the positive direction is shown in Fig. 5. The total energy was set slightly below the threshold value of 100 (99.995) in order to prevent the trajectory from escaping to infinity during the generation of the Poincaré surface of section shown in Fig. 5. The section itself is defined for $x=0$, with the index one saddle being at the origin, and trajectories are recorded as they pass through the plane only in the positive $x$ direction. In Fig. 5, only points on a single trajectory are plotted. The trajectory is initiated at the origin and run for $1.5 \times 10^6$ time units. The computation used a velocity Verlet integrator [8] with a time step of $10^{-4}$ units, and it conserved energy to $< 10^{-8}$ in absolute energy units (better than one part in $10^{10}$ in relative energy).

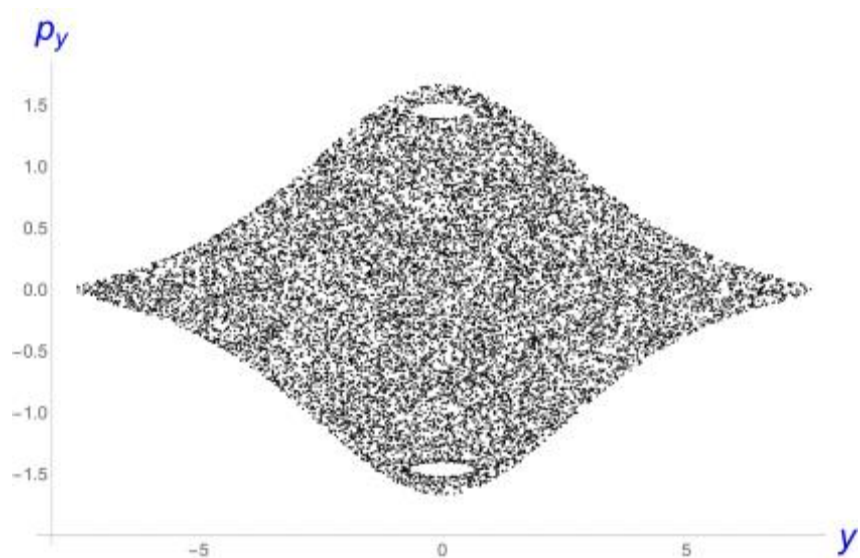

**Figure 5**. Poincaré surface of section for a trajectory initiated at $x = 0$, $y = 1$ and at a total energy of 99.995 on the double-Morse potential with parameter values $D_e = 100$, $k = 200$, $r_e = 1$, $b=5$, $m=1$. The trajectory was integrated for a total of $1.5 \times 10^6$ time units. The energy was kept slightly below that of the threshold (100 units), so that, the trajectory remained bounded.

Clearly, Fig. 5 is illustrative of chaotic dynamics, and so Moser's theorem rules out the existence of a second analytic integral in this case. However, note that the Poincaré section does not exhibit complete "global chaos". There are islands at the top and bottom of the Poincaré section corresponding to the intersection of a stable PO with the Poincaré section. This raises the question of the "statisticality" of roaming dynamics.

The apparently chaotic behavior revealed in Fig. 5 shows that the dynamics on this potential cannot be completely integrable. However, that does not mean that the dynamics at energies above the threshold value are necessarily "statistical." The reason is that there exists a "chemical timescale" set by the time for a trajectory to pass from one well to the other, or to cross the boundary for heading off to infinity. This timescale is typically ~10 time units or less, whereas it took more than $10^6$ time units to generate Fig. 5. On the relatively short, chemically relevant, timescale, the longer time chaotic behavior need not be (fully) realized. Moreover, it is likely that, at energies above threshold the system exhibits the phenomenon of chaotic scattering, where finite lifetime scattering trajectories coexist with a measure zero set of trapped (infinite lifetime) trajectories associated with the so-called "chaotic repellor" [9,10].

**Periodic Orbit Dividing Surfaces (PODS):** A geometrical object that is of great significance in chemical reaction dynamics is the notion of a *dividing surface* (DS). Mathematically, a dividing surface is simply a surface of one less dimension than the ambient space (i.e. it is said to have "codimension one"), having the property that all trajectories of interest pass through the DS. Of course, the key question is "what does the DS divide?" This is a problem dependent question, but for the double Morse potential the DSs will control the entrance and exit of trajectories to the three regions of the PES, which we have labelled A, B, and C in Fig. 2. The characterization of trajectories is a dynamical, or phase space notion and, consequently, the DS will in general be a 2*d* surface embedded in the 3*d* energy surface.

A detailed development of the construction of DSs for 2 DoF systems using a certain class (family) of unstable POs was given in a beautiful series of papers by Pollak, Pechukas, and Child in the late 1970's to early 1980's [11]. The resulting PO dividing surfaces or PODS, possess many desirable features for determining the rate of crossing of trajectories of this DS. In particular, the DS has the "no-recrossing" property. Mathematically, this means that the Hamiltonian vector field is transverse to the DS. Another property of the DS constructed in this manner is that the flux across the DS is minimal in the sense that perturbations to the DS have larger flux.

We now describe an algorithm for sampling points on the phase space DS defined by a general unstable PO for a 2 DoF Hamiltonian system with a Hamiltonian of the form given at Eq. 3. The procedure selects points on a 2*d* surface with fixed total energy $E$, where the PO forms the one dimensional (1*d*) boundary of the DS. The algorithm is as described in [12]:

1. Locate an unstable PO.
2. Project the PO into configuration space.
3. Choose points on that curve $(x_i, y_i)$ for $i = 1, \cdots, N$, where $N$ is the desired number of points. Points are spaced uniformly according to distance along the PO.
4. For each point $(x_i, y_i)$ determine $p_{x\,\max,i}$ by solving

$$H(x_i, y_i, p_x, 0) = \frac{p_x^2}{2m} + V(x_i, y_i) = E \qquad (4)$$

   for $p_x$. Note that solution of this equation requires $E - V(x_i, y_i) \geq 0$, and there will be two solutions, $\pm p_{x\,\max,i}$.

5. For each point $(x_i, y_i)$ choose points $p_{x_j}$ for $j = 1, \cdots, K$, with $p_{x_1} = 0$ and $p_{x_K} = p_{x\,\text{max},i}$ and solve the equation $H(x_i, y_i, p_x, p_y) = E$ to obtain $p_y$.

The next question concerns the geometrical structure of the DS. The DS sampled in this manner is a one parameter family of circles. The parameter defining the family is given by the distance along the projection of the PO onto the configuration space in Steps 1-3 in the algorithm above, and the momentum-space circles are given by the Hamiltonian:

$$\frac{p_x^2}{2m} + \frac{p_y^2}{2m} = E - V(x_i, y_i) \tag{5}$$

More information about the geometry of this one parameter family of circles is obtained from the nature of the projection of the PO into configuration space. In the work of Pollak, Pechukas, and Child [11], the configuration space projections of the POs were arcs (diffeomorphic to an interval) where a configuration space point on the projection of the PO moves back and forth along the arc. Necessarily, the endpoints of the arc are turning points with $p_x = p_y = 0$, where the circles defined by Eq. 5 shrink to points. This implies that the one parameter family of circles defines a 2-sphere. We will not discuss this type of PO in more detail, since, these are not the types of POs that we will consider in this paper.

Rather, the POs that we will study have the property that their projection into configuration space is either a simple closed curve (i.e. diffeomorphic to a circle) or it is a closed curve with one point of self-intersection (like a figure eight). In both cases the DS is a two-torus, $S^1 \times S^1$. The circles defined by Eq. 5 "sweep out" the two-torus as they move along the closed curve projection of the PO in configuration space. Each point along this projection defines a circle in the momentum space. On this circle, there are precisely two momentum values, of equal magnitude and opposite sign corresponding to two POs that evolve in the longitudinal direction on the surface of the torus in opposite senses. These two POs serve as boundaries to two halves of the DS.

The 2$d$ DS can be parametrized by the arc length along the configuration space projection of the PO and one component of the momentum vector (the other component is fixed by energy conservation). Towards this end, it is natural to decompose the momentum vector into a component tangent to the PO and a component normal to the PO, denoted $p_{tangent}$ and $p_{normal}$, respectively. We will use $p_{normal}$ in our parametrization of the DS. In this case, positive $p_{normal}$ corresponds to the outgoing component of the DS and negative $p_{normal}$ corresponds to the incoming component of the DS.

We remark that unstable POs do not have high enough dimension to construct a DS for systems with more than 2 DoF. The appropriate construction requires use of a normally hyperbolic invariant manifold (NHIM) [3].

**Unstable Periodic Orbits:** Three classes (families) of unstable POs have been found for this PES, labeled as Type 1, Type 2, and Type 3. These POs will be used to construct PODSs in order to study trajectories that visit the regions A, B, and C shown in Fig. 2. The POs are shown for various total energies above the threshold, and two values of $b$, in Figures 6-11.

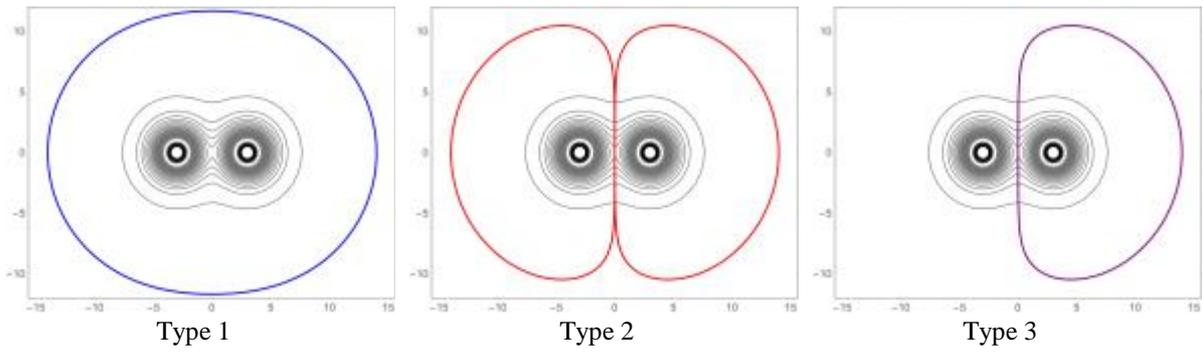

**Figure 6**. Unstable POs for $E = 100.001$ and $b = 3$.

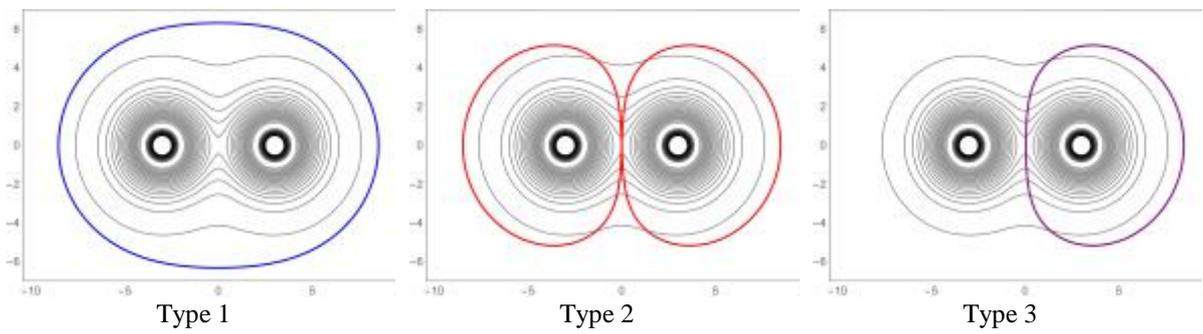

**Figure 7**. Unstable POs for $E = 101$ and $b = 3$.

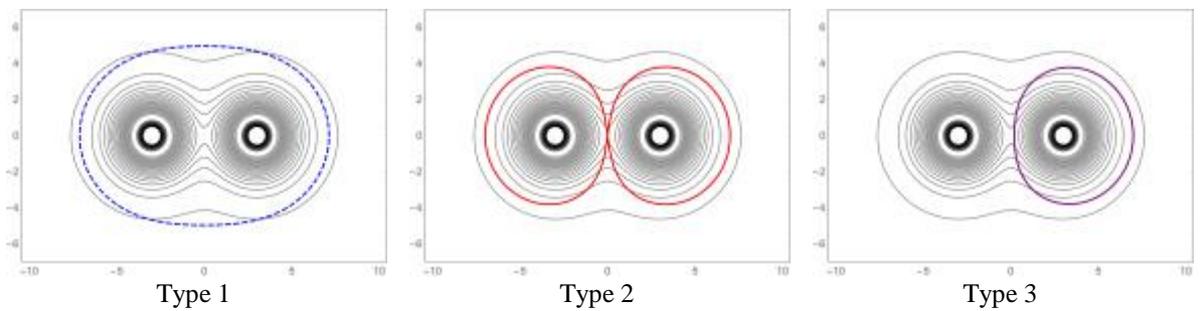

**Figure 8.** Unstable POs for $E = 105$ and $b = 3$.

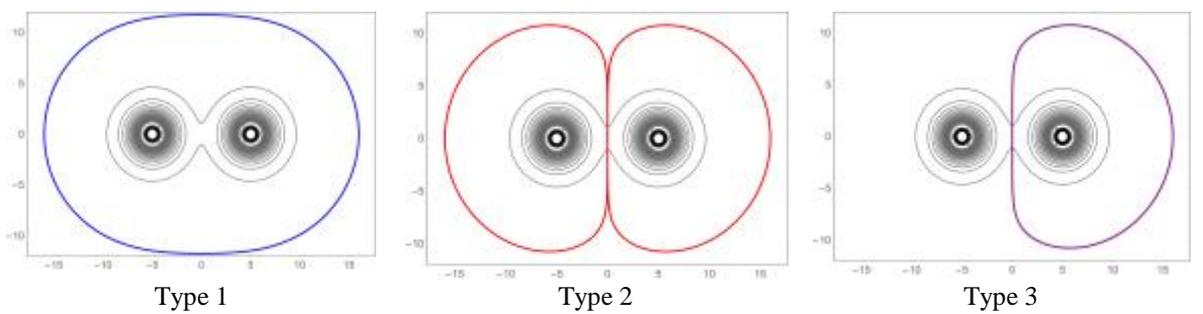

**Figure 9**. Unstable POs for $E = 100.001$ and $b = 5$.

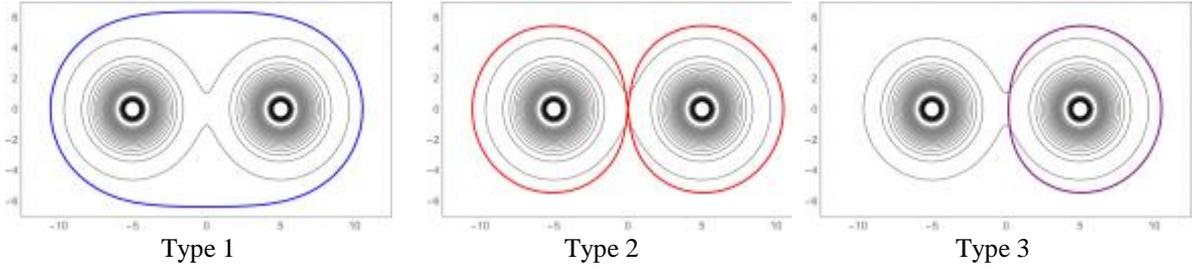

Type 1    Type 2    Type 3

**Figure 10**. Unstable POs for *E* = 101 and *b* = 5.

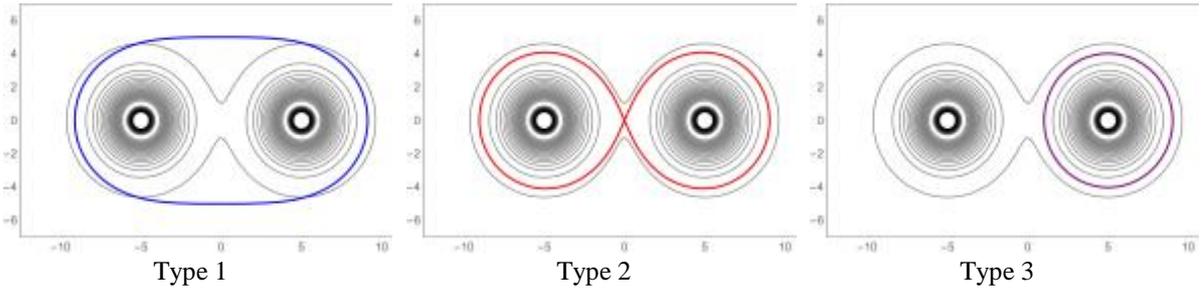

Type 1    Type 2    Type 3

**Figure 11**. Unstable POs for *E* = 105 and *b* = 5.

As noted in our discussion of PODs, since the Hamiltonian is quadratic in the momentum, the POs occur in pairs that rotate in opposite senses. Therefore, the Type 1 and Type 2 POs each come in pairs, corresponding to opposite senses of rotation. Type 3 also has two senses of rotation, but additionally exists on the left and right halves of the potential, and so comes as a set of four. For the parameter values that we considered, Type 3 is everywhere inside Type 2 in configuration space, and Type 2 is everywhere inside Type 1. It is also apparent that the POs contract as the energy increases. Indeed, the Type 1 PO is *analogous* to a centrifugal barrier that would be present in a spherically symmetric potential (e.g. a Kepler problem), and the contraction of this PO with increasing energy is analogous to the similar shift seen in a centrifugal barrier. This is important because we hypothesize that the Type 1 PO defines the configuration-space projection boundary within which roaming can occur. This conjecture is based on the empirical fact that of the many thousand trajectories we have run on this potential, we have never seen one start inside this PO, cross it, and then do anything but head off to infinity. This point is further discussed in the Summary and Conclusions.

**Inter-Well Transport: The Role of Type 1 and Type 2 Periodic Orbits:** In this section we show a variety of trajectories in configuration space with different initial conditions for *b* = 5 and *E* = 100.001. Note that the gap between these two POs is very small at extreme values of $x$.

In Figs. 12 and 13, it is clear that inter-well transport is controlled by the Type 1 and Type 2 POs. Trajectories leave a well, enter the flat region of the PES and are turned back to one well or the other by the Type 1 PO. We will consider this further by initializing trajectories on the DS constructed from these POs, as described earlier.

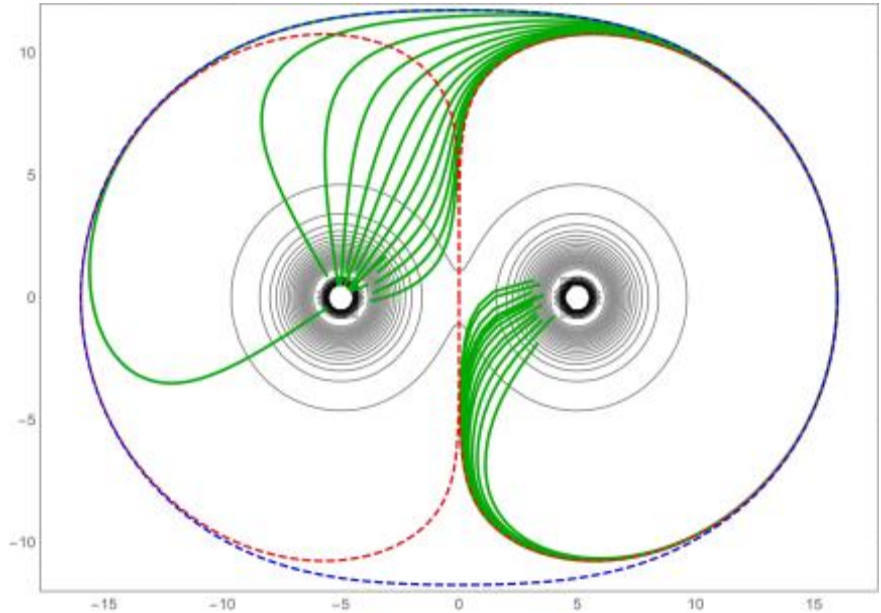

**Figure 12.** Several trajectories were started on the $x = 0$ line, at various values of $y$ from 6 to near the value corresponding to the Type 1 PO. They were integrated forward and backward in time and the relative magnitudes of $p_x$ and $p_y$ adjusted to show transport between the two potential wells. The value of $b$ is 5, and the energy 100.001. Type 1 (blue) and Type 2 (red) POs are also shown.

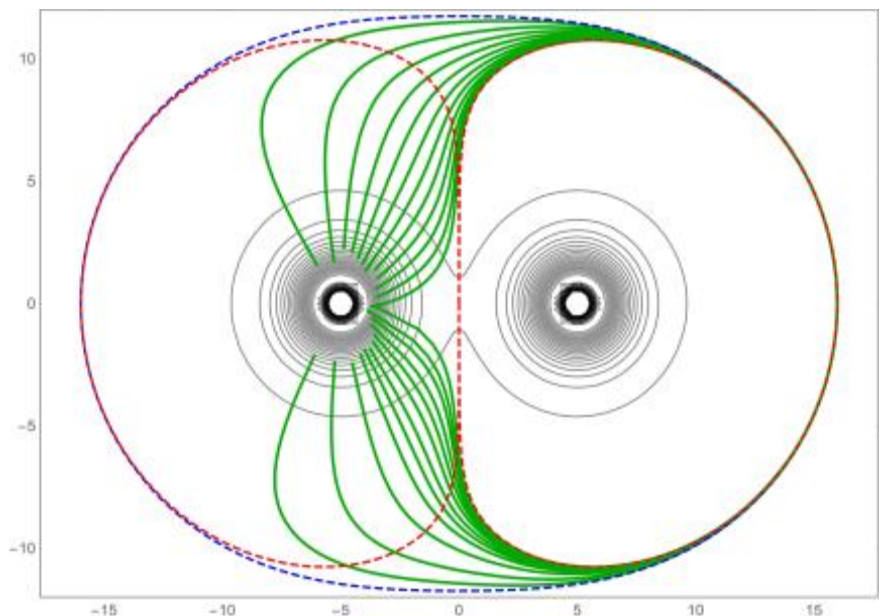

**Figure 13.** Several trajectories were started on the $x = 0$ line, at various values of $y$ from 6 to near the value corresponding to the Type 1 PO. They were integrated forward and backward in time and the relative magnitudes of $p_x$ and $p_y$ adjusted to show termini in the left-hand well. The value of $b$ is 5, and the energy 100.001. Type 1 (blue) and Type 2 (red) POs are also shown.

**Computation and Sampling of PODS for Type 1, Type 2, and Type 3 Periodic Orbits:**
We now construct DSs associated with each of the POs. We will then choose initial conditions on the DS and evolve them forward and backward in time to determine their origin and fate. This will enable us to determine how a particular PO influences the fate of a trajectory according to its initial condition. The origin and fate will be denoted by X → Y, where origin X and fate Y are A, B, or C. We will denote the different origins and fates by color where blue is A→C, purple is B→C, green is C→C, yellow is A→A, grey is B→B, and red is A→B.

Recall that the $2d$ DS is parametrized by arc length along the PO and one component of momentum, either $p_{tangent}$ or $p_{normal}$. We will provide parametrizations where both components are utilized. Recall also that the components of momentum occur with equal magnitude and opposite signs because of the particular form of the kinetic energy. The component of momentum that we select will always be positive, which indicates that in positive time the trajectories evolve "outwards".

Type 1 PO
The arc of the Type 1 PO which was sampled is shown in bold in Fig. 14 for $E = 101$ and $b = 5$. Regions A–C have the same meaning as they did in Fig. 2. The red dot shows the starting point of the sampling.

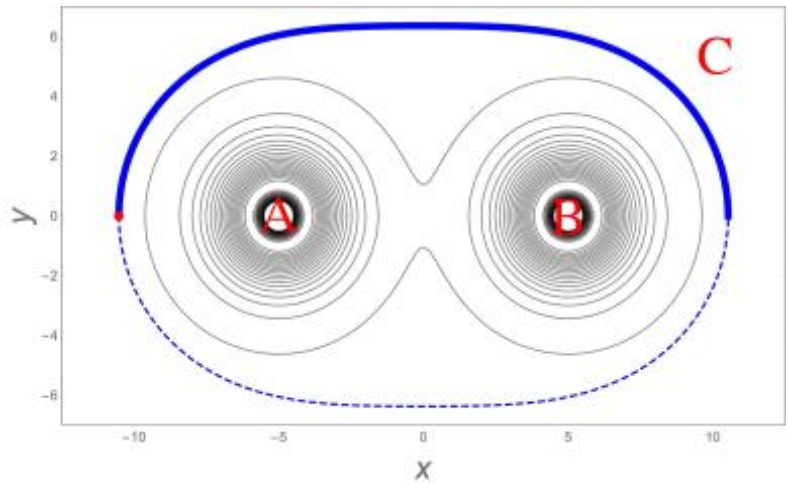

**Figure 14.** Arc of the Type 1 PO (shown in bold) sampled for the DS for $E = 101$ and $b = 5$. The configuration space arc length of the PO was calculated, starting at the red dot; it forms the basis of the abscissæ in Figs. 15 and 16.

One thousand points along the arc shown in Fig. 14 were sampled. For each point, the constant-length momentum vector was rotated from tangent to the PO to normal to the PO in 1000 steps. A trajectory was then run both forwards and backwards in time with those initial conditions until it reached one of the defined areas A–C shown in Fig. 14. The resulting fate maps are shown in Figs. 15 and 16.

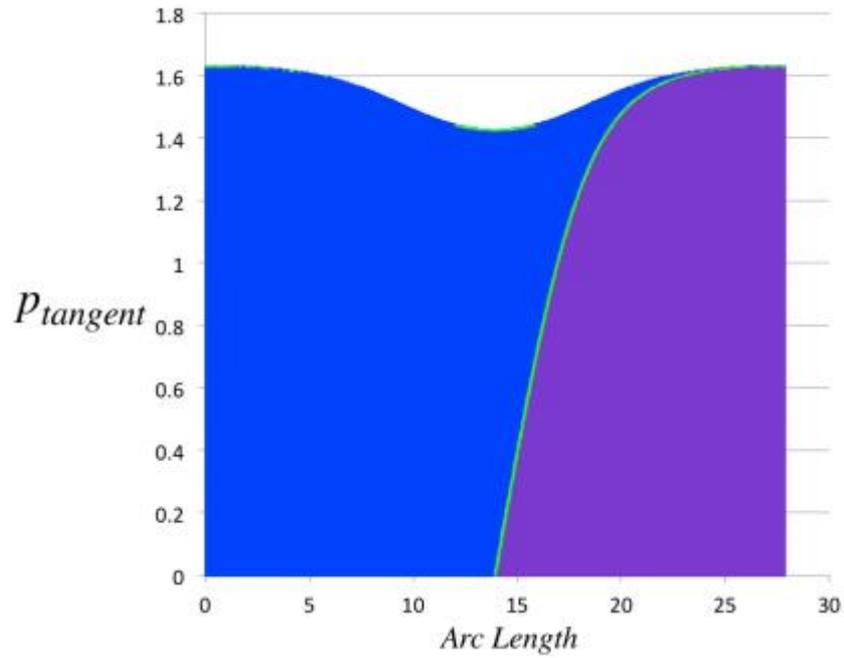

**Figure 15**. Fate map for the trajectories sampled along the arc of the Type 1 PO shown in Fig. 14. The color coding is: A→C (blue), B→C (purple), and C→C (green).

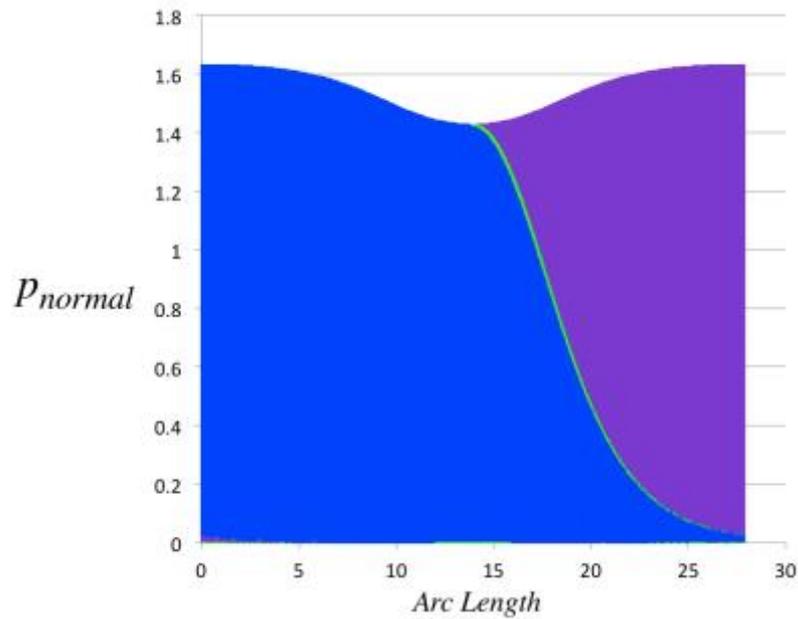

**Figure 16**. Fate map for the trajectories sampled along the arc of the Type 1 PO shown in Fig. 14. Color coding is the same as before: A→C (blue), B→C (purple), and C→C (green).

These results show that the Type 1 PO controls dissociation in the sense that all trajectories that pass through the DS constructed from this PO become unbounded (in positive time). Moreover, the trajectories that pass through the DS have one of the three possible origins, A, B, or C (as determined up to numerical precision).

Type 2 PO

Next, we will construct the DS for the Type 2 PO. This is the DS that should reveal the nature of inter-well transport.

In the results that follow, Figures 17-24, we sample momenta from 0 to $+p_{tangent}$ over the entire PO. Although there is redundancy in this information, the requirement for color complementarity between left and right halves of the plots, serves as a check against integration error in the calculations.

Fate Maps of the Type 2 PODS as a Function of *b*.

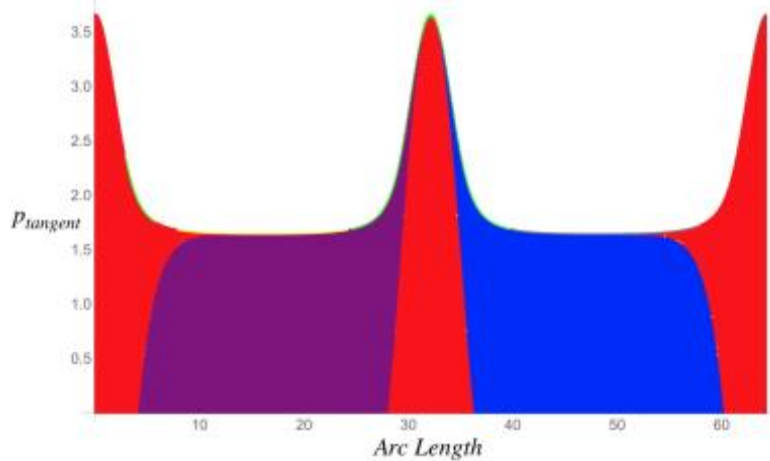

**Figure 17**. Fate map (tangent) for a sample along the entire arc length of the Type 2 PO with $b = 4$.

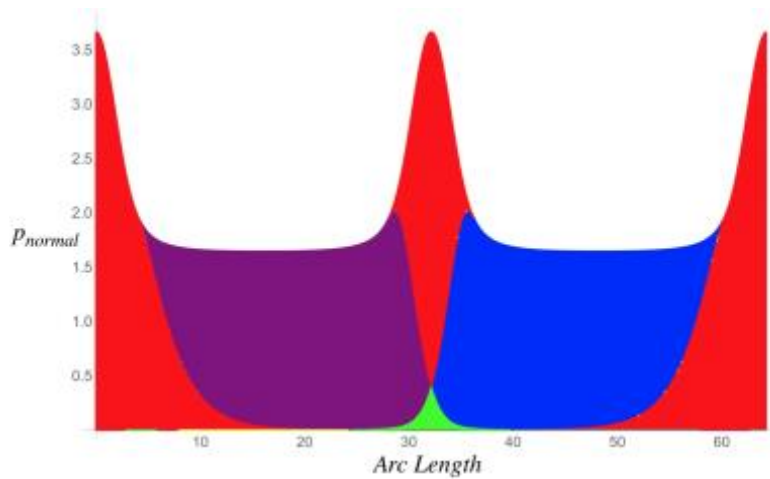

**Figure 18**. Fate map (normal) for a sample along the entire arc length of the Type 2 PO with $b = 4$.

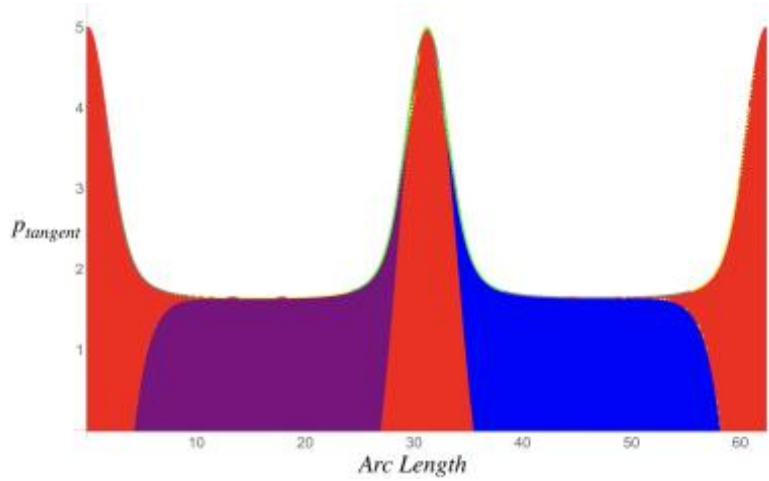

**Figure 19**. Fate map (tangent) for a sample along the entire arc length of the Type 2 PO with $b = 3.5$.

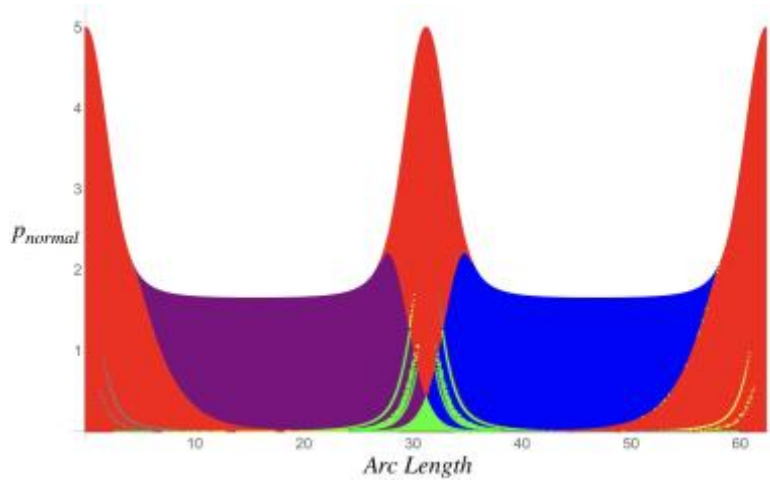

**Figure 20**. Fate map (normal) for a sample along the entire arc length of the Type 2 PO with $b = 3.5$.

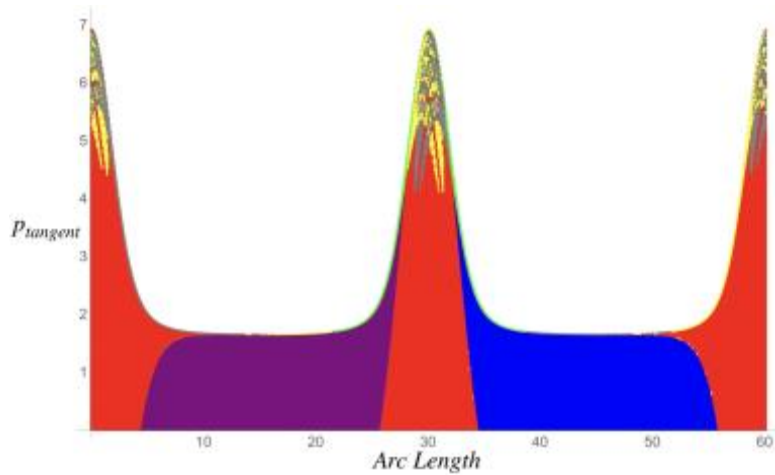

**Figure 21**. Fate map (tangent) for a sample along the entire arc length of the Type 2 PO with $b = 3$.

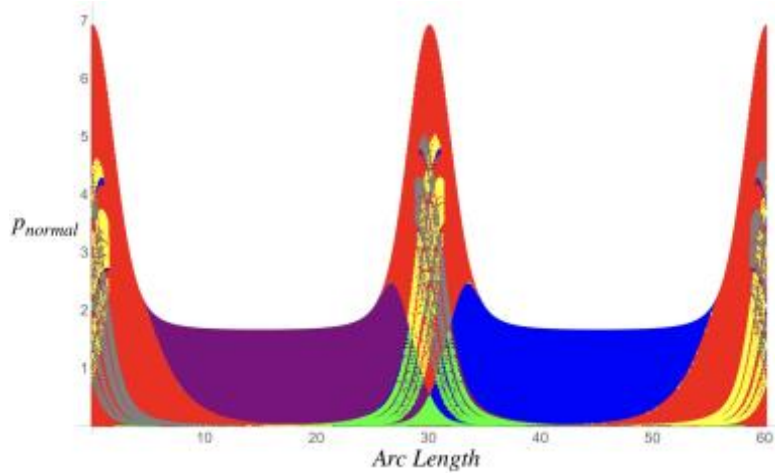

**Figure 22**. Fate map (normal) for a sample along the entire arc length of the Type 2 PO with $b = 3$.

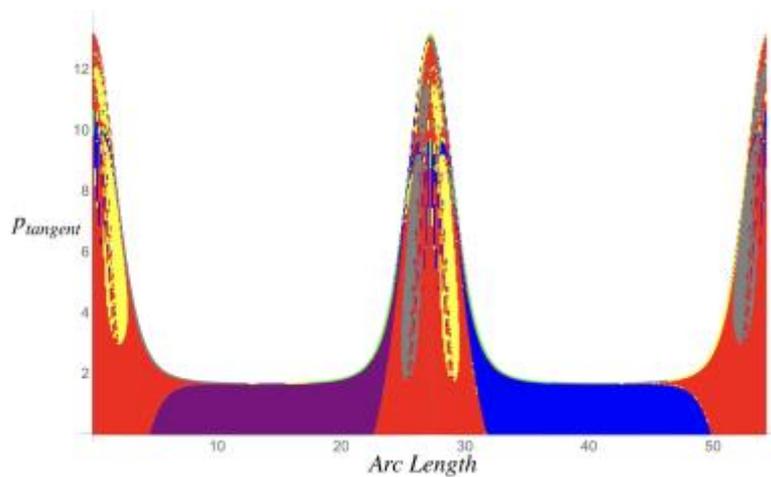

**Figure 23**. Fate map (tangent) for a sample along the entire arc length of the Type 2 PO with $b = 2$.

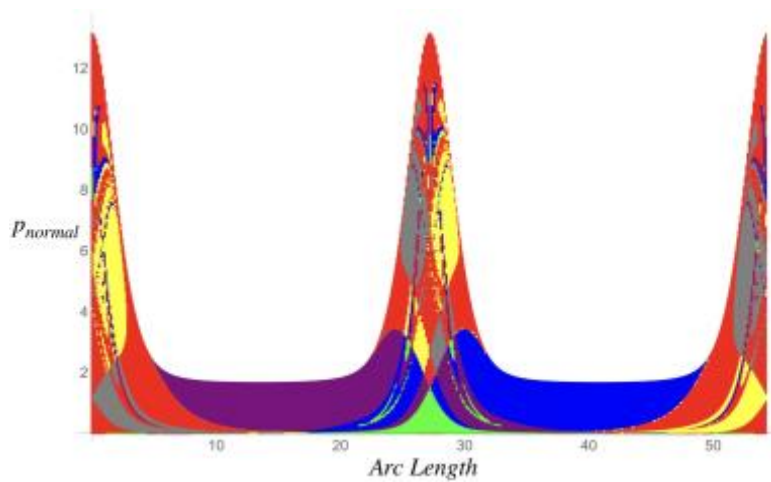

**Figure 24**. Fate map (normal) for a sample along the entire arc length of the Type 2 PO with $b = 2$.

Qualitatively new dynamical phenomena start to appear at $b \sim 3.5$ and grow in complexity as $b$ decreases in value. In order to more fully investigate this, we compute individual trajectories for $b = 3$, with initial conditions being near the centers of Figs. 21 and 22. The locations of the specific initial conditions are shown in Fig. 25.

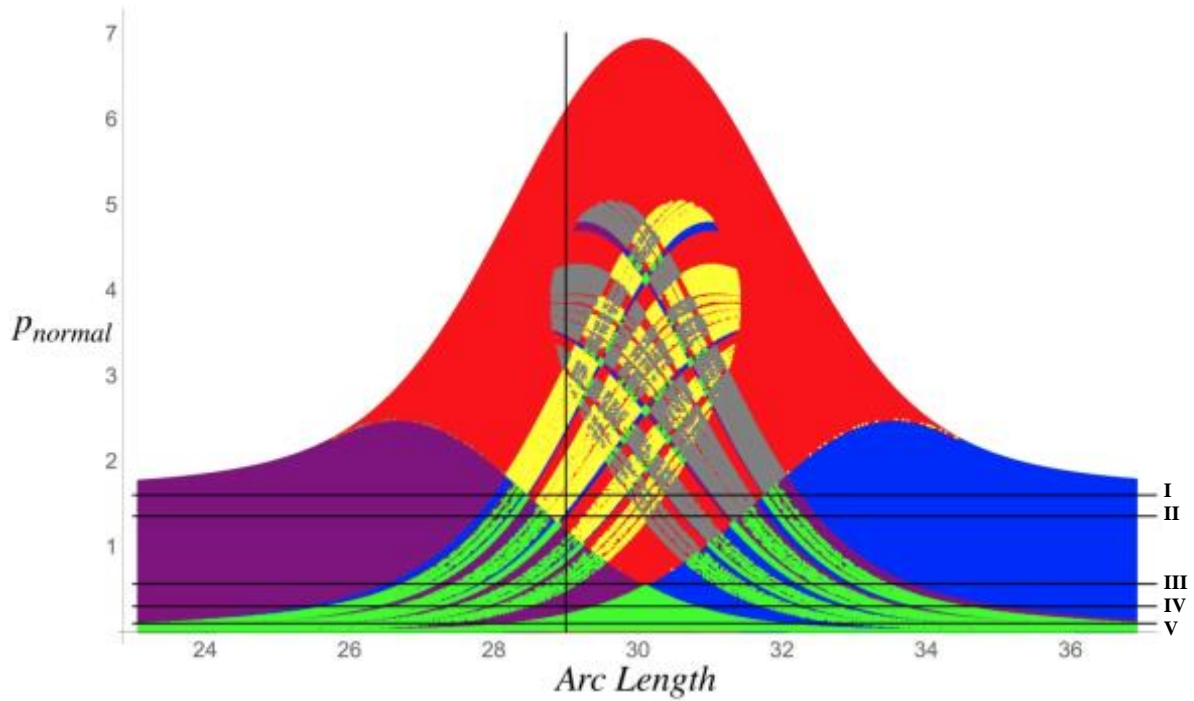

**Figure 25**. Scale expansion of Fig. 22 with lines whose intersections show the initial conditions for the individual trajectories shown in Fig. 26.

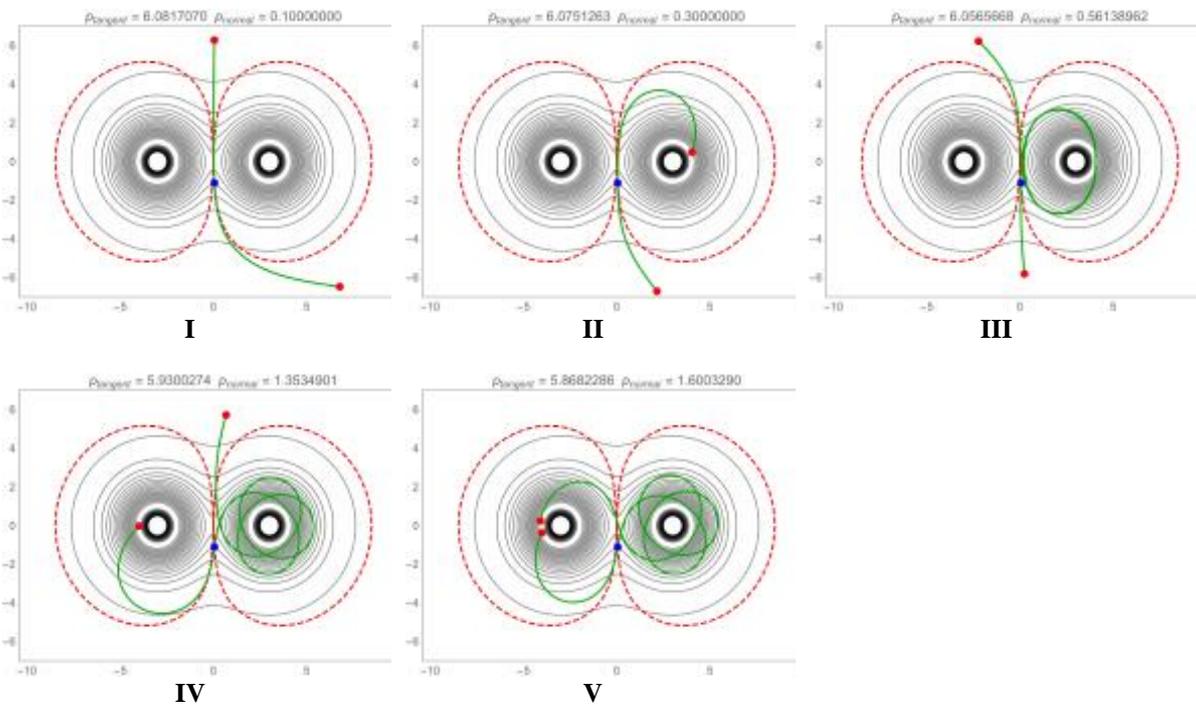

**Figure 26**. Individual trajectories, labeled according to their initial conditions, as shown in Fig. 25.

The interesting-looking dynamical structures appearing on the Type 2 PO dividing surfaces seem to arise from temporarily trapped trajectories in either one well or the other of the potential. The criterion that we use for terminating a trajectory in a well is that it should approach the center of that well to within a distance of ≤ 1 unit, which is where the local minimum of the potential occurs. Trajectories such as Fig. 26C do not quite meet this criterion and so are allowed to continue, leading, as in Fig. 26C, to dissociation in both directions. Had we chosen a looser criterion for termination, much of the structure appearing in the DSs would have disappeared. Moreover, the reason that the structure appears only for low values of *b* is that the total available momentum in the region of the index one saddle increases as *b* decreases, and one needs a certain amount of angular momentum° to trap trajectories in a well. For values of *b* > 3.5 there is not enough angular momentum available to lead to trapping. Nonetheless, we do observe inter-well transport in Fig. 25 near the junctions of the large blue (A→C) and large red (A→B) fate regions and near the junctions of the large purple (B→C) and large red (A→B) fate regions. A close inspection shows that the former junction has a line of yellow (A→A) dots while the latter junction has a line of gray (B→B) dots along it. In Fig. 27, we show what is going on at the red-yellow-blue interface in Fig. 25. The trajectory initial conditions for Fig. 27 cannot easily be shown on Fig. 26, because the differences in initial momenta are so small that the horizontal lines depicting the values of $p_{normal}$ could not be resolved on the diagram. Overall, the apparently fractal complexity seen in Fig. 25 is the signature of chaotic scattering above threshold [9].

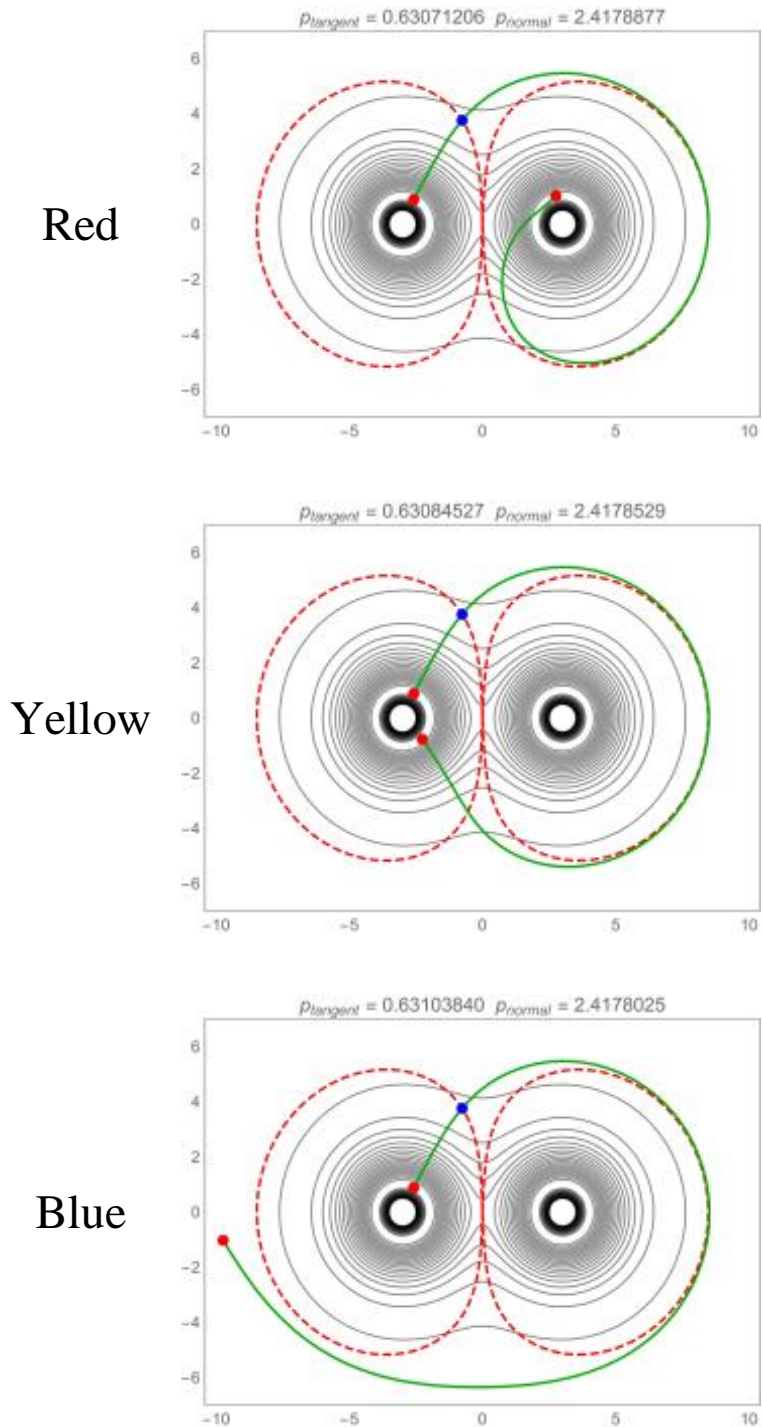

**Figure 27**. Inter-well transport occurring at the red-yellow-blue interface in Fig. 25.

Type 3 PO

The arc of this PO which was sampled is shown in bold in Fig. 28. The red dot shows the starting point of the sampling. Results are presented in Figs. 29 and 30.

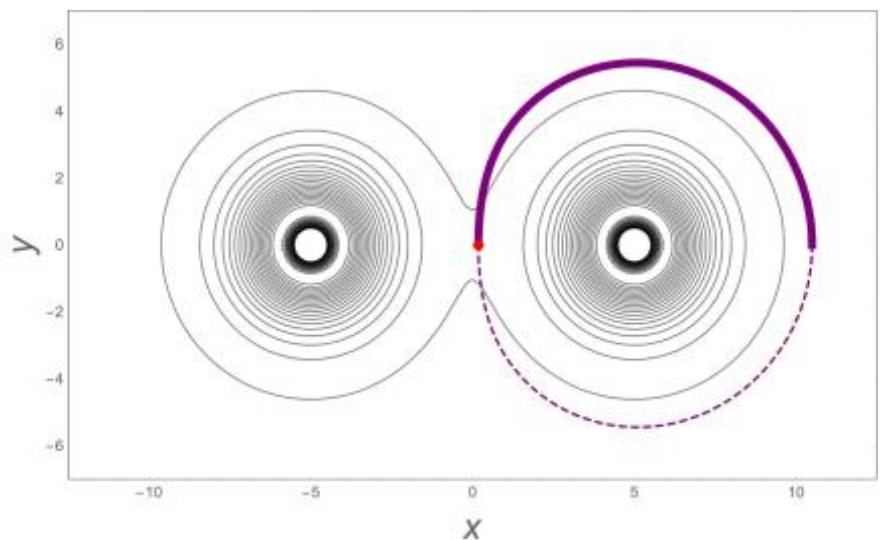

**Figure 28.** Arc of the Type 3 PO (shown in bold) sampled for the DS. The configuration space arc length of the PO was calculated, starting at the red dot; it forms the basis of the abscissæ in Figs. 29 and 30.

---

° In this paper, we use the term angular momentum (AM) in this context to be the AM of the trajectory with respect to the center of the well it is orbiting. If the trajectory is circling the right well, as in Fig. 26C-D, the origin is taken to be (*b*,0), the center of the right well when computing the AM. It should be stressed that the potential is not cylindrically symmetric about either well, however, and the AM as defined is not a conserved quantity on this PES.

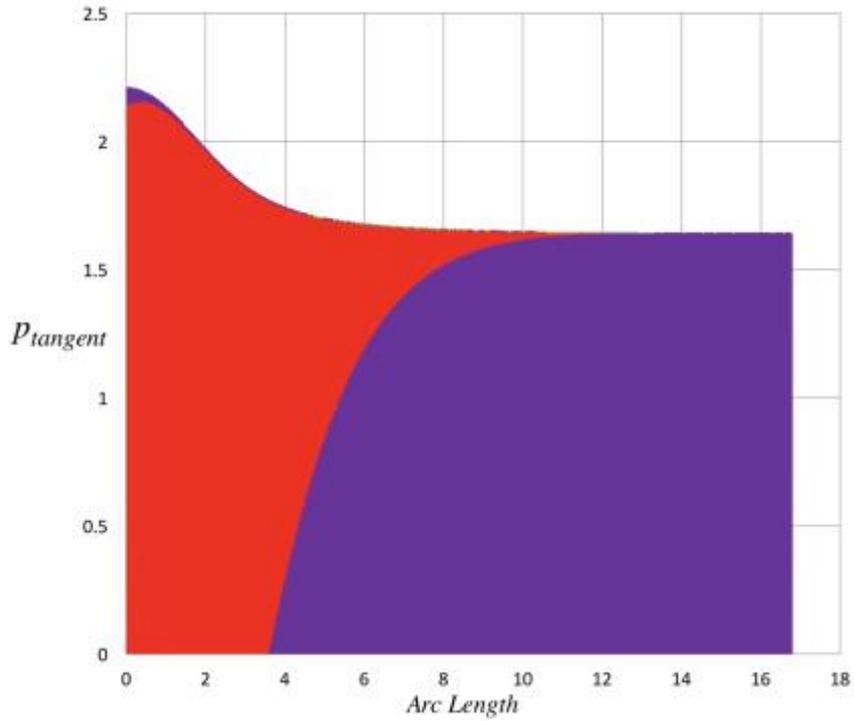

**Figure 29**. Fate map for the trajectories sampled along the arc of Type 3 PO shown in Fig. 28. Color coding is the same as before: A→B (red), and B→C (purple).

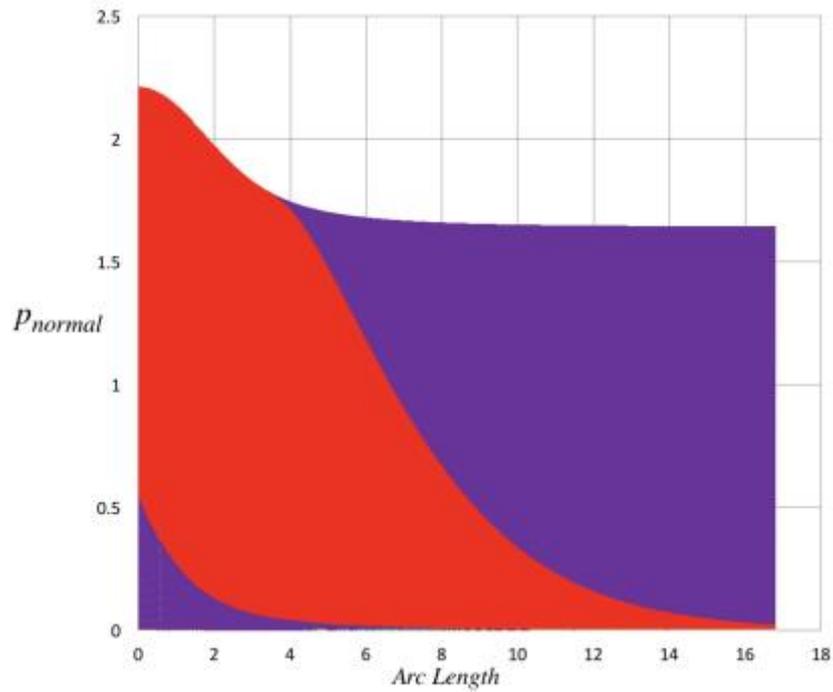

**Figure 30**. Fate map for the trajectories sampled along the arc of Type 3 PO shown in Fig. 28. Color coding is the same as before: A→B (red), and B→C (purple).

**Summary and Conclusions:** At this point we summarize our main findings.

1. The dynamics on the Double-Morse potential, at least at energies below the threshold, are chaotic (Fig. 5) and therefore not completely integrable.
2. By virtue of Assertion 1, one can conclude that angular momentum is not conserved for dynamics on this potential, and that there exists no other conserved quantity (except energy). The role played by angular momentum for the Double-Morse potential will be considered in another publication.
3. There are three classes (families) of unstable PO whose associated DSs appear to control the dynamics on this potential at energies above threshold (Figs. 6-11).
4. The DS separating trapped trajectories from escaping trajectories within one well appears to be derived from the Type 3 PO. There do exist trajectories on the double-Morse potential that look similar to roaming trajectories in other systems (e.g., Figs. 4, 12, 13). However, there is no "roaming saddle" on this potential.
6. The gross separation of dissociating trajectories from all others appears to be controlled by the DS associated with the Type 1 PO (Figs. 14-16). It appears that this PO occupies a role similar to that associated with a centrifugal barrier on a cylindrically symmetrical potential. Once a trajectory is outside the Type 1 PO (in configuration space) and moving to larger values of *r*, it never returns. *We conjecture that once a trajectory is outside any of the POs (in configuration space) and moving to larger values of r, that PO (or its associated stable and unstable manifolds) no longer controls the dynamics of the trajectory.*
7. The implications of the conjecture in Point 6 can be brought into focus by considering the trajectories in Fig. 27. Once the trajectory crosses the Type 2 PODS at the blue dot, moving outwards, its fate, according to our conjecture, is determined entirely by the Type 1 PO. All that PO can do is control whether the trajectory leads to dissociation or not. In the Blue trajectory of Fig. 27, the condition for crossing the Type 1 PODS is satisfied, whereas in the Red and Yellow trajectories in Fig. 27, it is not. However, exactly what that Type 1 PO crossing condition is remains elusive. We have observed that the angular momentum (defined with respect to one of the two wells) to be a dynamically interesting quantity, which we explore in another paper [13]. This quantity provides a qualitative measure for dividing A→A and B→B trajectories from A→C and B→C trajectories. Nevertheless, criteria based on angular momentum are only qualitative, and the lack of angular momentum conservation or anything like it means that the quantity does not provide predictive capability. Computation of stable and unstable manifolds associated with Type 1 and Type 2 POs will likely be necessary to fully elucidate the dynamics.
8. If a trajectory fails to meet the criterion for crossing the Type 1 PODS it will necessarily turn around and the coordinate *r* will begin to decrease. At this point, the Type 2 PO can regain control of its fate. In the Yellow trajectory of Fig. 27, the criterion for recrossing the original Type 2 PODS (recrossing, because the trajectory crossed it once on the way out) is not satisfied whereas for the Red trajectory of Fig. 27, it is. We say "original Type 2 PODS" because there is a subtlety here. As we have remarked before all of these POs come in counter-rotating pairs. The one that we sampled to get the DSs and the trajectories in Fig. 27 has the particular sense of rotation shown by the arrows in Fig. 31 (starting at the blue arrow and ending at the red one).

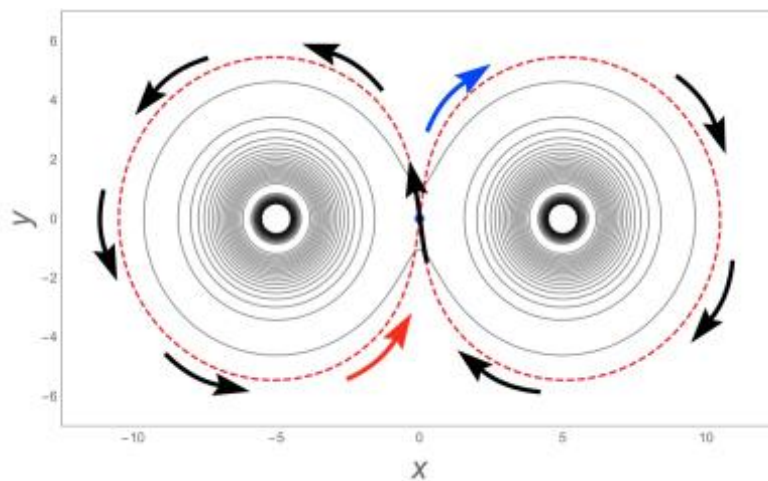

**Figure 31**. Sense of rotation of the PO sampled to create the DS in trajectories in Fig. 27.

If one thinks about the direction of motion of the green trajectories in Fig. 27, moving outwards from the left-hand well, one sees that the sense of motion is the same as that for the right-hand half of the PO in Fig. 31, but opposite to that in the left-hand half. Hence, it seems that the DS controlling behavior of the Yellow trajectory in the region {$x<0$, $y<0$} is actually the oppositely rotating one to the PO shown in Fig. 31. In other words, in order to completely understand the behavior of trajectories on this potential, one may have to consider individually the DSs associated with each of the counter-rotating pairs of POs.

9. The last point is of some relevance for understanding the nature of roaming dynamics. The trajectories apparently demonstrating roaming behavior (e.g., Figs. 4, 12, 13) are found *not* to belong to distinct parts of the Type 2 PODS from those corresponding to "conventional" reactions. Although dynamically distinct events corresponding to several circuits around the two wells must exist, our efforts to display them on a DS convince us that they constitute an entirely negligible fraction of the overall system dynamics.

**Acknowledgements:** The authors are grateful for useful communications with Alexey Borisov related to the two center problem. SW acknowledges the support of ONR Grant No. N00014-01-1-0769. SW and BKC acknowledge the support of EPSRC Grant No. EP/P021123/1.